\documentclass[twocolumn,amsmath,amssymb,nofootinbib,eqsecnum,tighten,prl]{revtex4-1}

\usepackage{graphicx}
\usepackage{dcolumn} 
\usepackage{bm, bbm}      
\usepackage{curves}
\usepackage{epic}
\usepackage{wasysym}
\usepackage{color}
\usepackage{epsf, times}
\usepackage{subfigure}
\usepackage{amsthm}

\def\openone{\leavevmode\hbox{\small1\kern-4.2pt\normalsize1}}

\newcommand{\ssection}[1]{{\em #1:}}

\newcommand{\beq}{\begin{equation}}
\newcommand{\eeq}{\end{equation}}
\newcommand{\bei}{\begin{itemize}}
\newcommand{\eei}{\end{itemize}}
\newcommand{\bea}{\begin{eqnarray}}
\newcommand{\eea}{\end{eqnarray}}
\newcommand{\bfig}{\begin{figure}}
\newcommand{\efig}{\end{figure}}

\begin{document}


\title{
Dipolar order by disorder in the classical Heisenberg antiferromagnet on the kagome lattice
      }

\author{
Gia-Wei Chern$^{1,2,3}$
}
\author{
R. Moessner$^1$
       }
\affiliation{
$^1$
Max-Planck-Institut f\"ur Physik komplexer Systeme,
Dresden, Germany
            }           
 \affiliation{
$^2$
           Theoretical Division,  Los Alamos National Laboratory, Los Alamos, New Mexico, USA}
\affiliation{
$^3$
Department of Physics, University of Wisconsin, Madison, Wisconsin, USA
}

\date{\today}

\begin{abstract}
Ever since the experiments which founded the field of highly frustrated magnetism, the kagome Heisenberg antiferromagnet has been the archetypical setting for the study of fluctuation induced exotic ordering. To this day the nature of its classical low-temperature state has remained a mystery: the non-linear nature of the fluctuations around the exponentially numerous harmonically degenerate ground states has not permitted a controlled theory, while its complex energy landscape has precluded numerical simulations at low temperature, $T$. Here we present an efficient Monte Carlo algorithm which removes the latter obstacle. Our simulations detect a low-temperature regime in which correlations asymptote to a remarkably small value as $T \to 0$. Feeding these results into an effective model and analyzing the results in the framework of an appropriate field theory implies the presence of long-range dipolar spin order with a tripled unit cell. 
\end{abstract}

\maketitle
%
%

The first experiments on the `kagome bilayer' SCGO~\cite{ObradSCGO88,ramespcoo} triggered a wave of interest in kagome antiferromagnets in particular, and frustrated systems in general. 
A cluster of early seminal theoretical papers~\cite{CHS,HBK,HRkag,CCRkag,ZEkag} established kagome magnets as model systems for novel ordering phenomena, discussing in particular spin liquidity, partial order, disorder-free glassiness and order by disorder. The excitement persists, not least in the quantum realm  \cite{balentsnature}, where there has been much recent progress in understanding the ground state for $S=1/2$ \cite{HuseWhite}.

\begin{figure}[ht]
\begin{center}
\includegraphics[width=0.8\columnwidth]{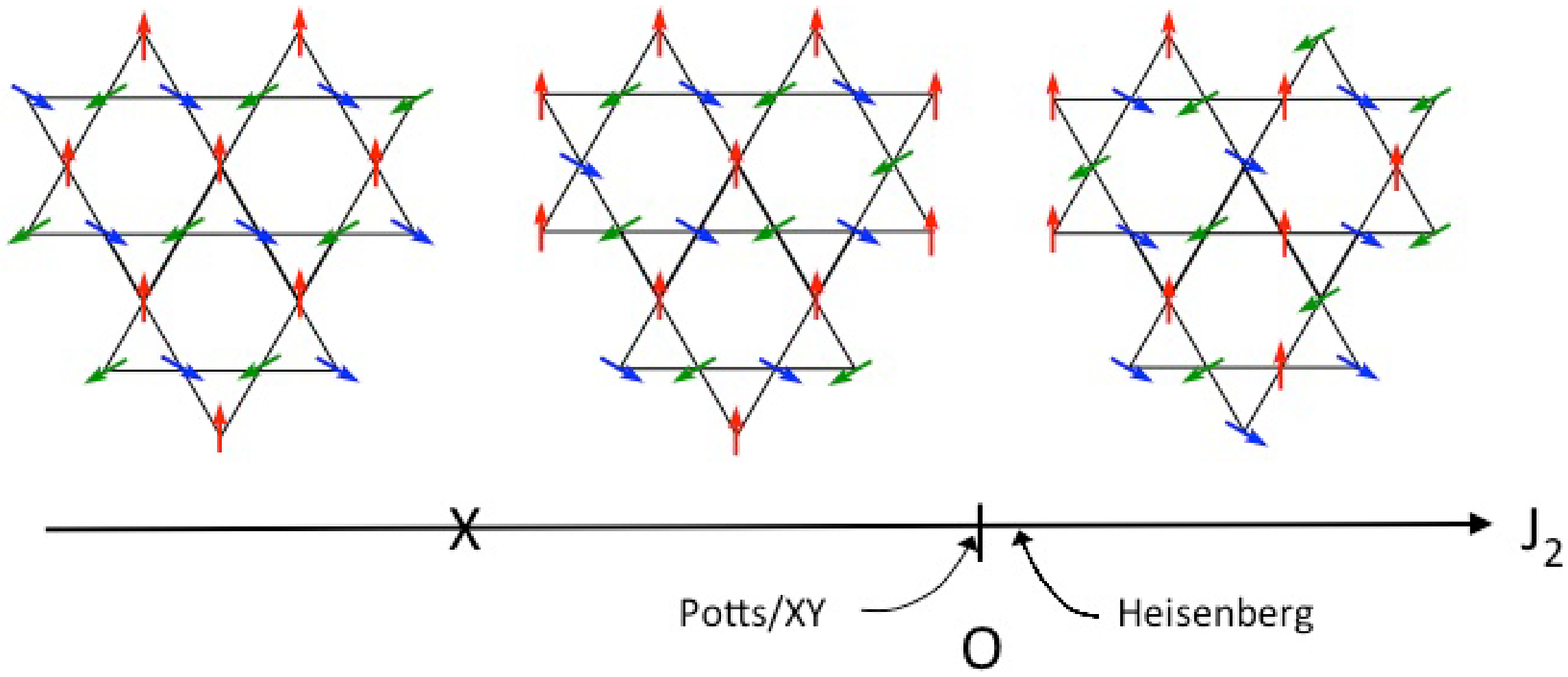}
\includegraphics[width=0.8\columnwidth]{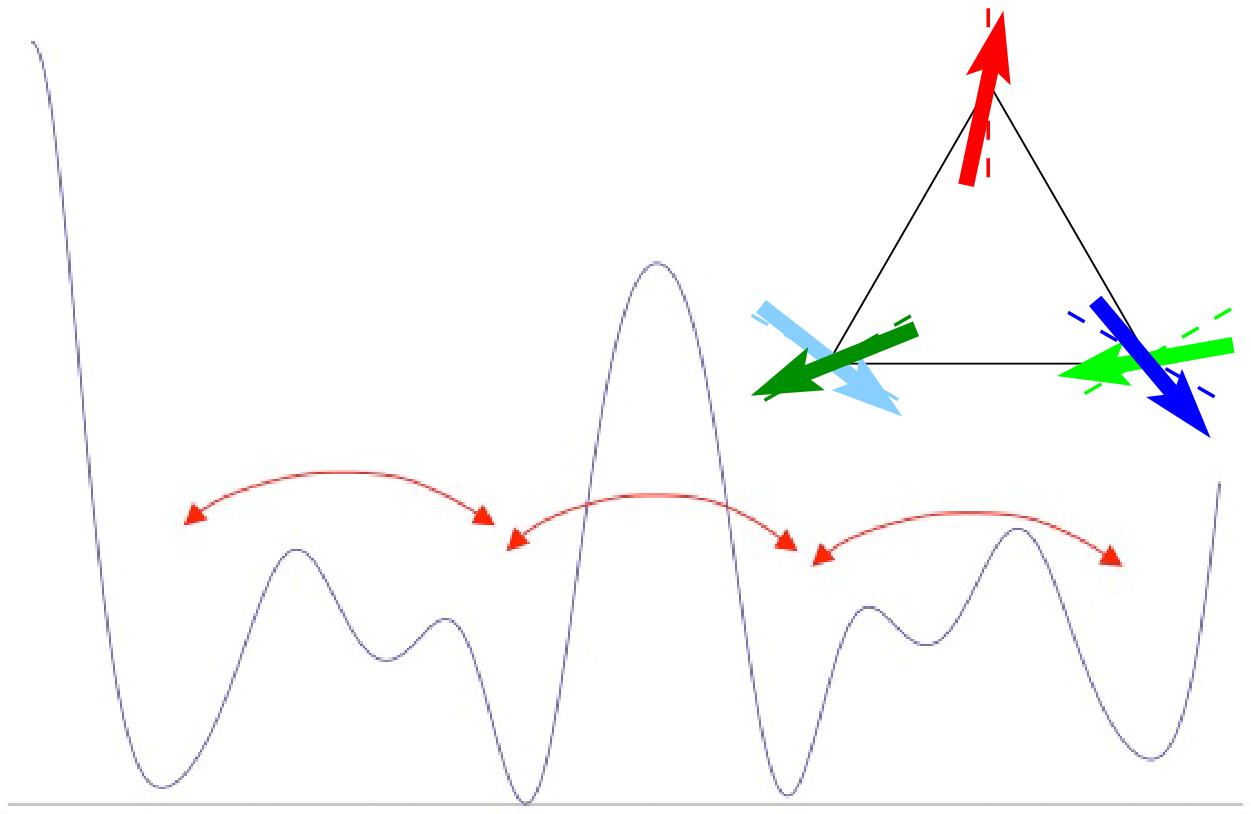}
\end{center}
\caption{
\label{fig:kagbasics} 
Top: candidate phases of the kagome antiferromagnet (from left to right: $q=0$; 
Kosterlitz-Thouless phase; $q=\sqrt{3}\times\sqrt{3}$) in the phase diagram of an extended Potts model as 
defined in the text. $J_2=0$ corresponds to the pure Potts model on the kagome lattice, the unweighted ensemble 
of coplanar Heisenberg ground states. 
Bottom: cartoon of the  effective entropy landscape for ground states--soft fluctuations yield minima at co-planar states. 
Our algorithm surmounts the entropic free energy barriers, exploring many   coplanar states. The resulting ensemble average leads to 
a small but nonzero spin order. Inset: transitions between harmonically equivalent configurations--spins on loops of alternating color are rotated such that the angle a spin makes with its local Potts axis is not changed as the dark green/blue spins are exchanged for the light ones.}
\end{figure}

Remarkably, for the {\em classical} kagome Heisenberg magnet, the nature of low-temperature phase has not been established, despite the deceptive simplicity of its Hamiltonian, encoding only nearest-neighbour interactions of strength $J>0$ between classical unit-length spins ${\bf S}_i$: $H=J\sum_{\langle ij\rangle}{\bf S}_i\cdot{\bf S}_j$.  This happens because  classical Heisenberg spins do not lend themselves at all to the numerical  methods applied to $S = 1/2$ (e.g.~\cite{sindzin,lalh,HuseWhite}), while classical Monte Carlo (e.g. \cite{CHS,Reim,HRkag,cepas}) has not been able to sample the different local free energy minima separated by entropic barriers -- for the best effort yet, see Ref. \cite{Zhito}. At the same time, analytical approaches have managed to develop different possible scenarios, described below, without being able to choose between them~\cite{HRkag,Henley}.

Here, we present a Monte Carlo algorithm which enables us to simulate systems containing over 2000 spins down to a low-temperature regime, where we find that correlations no longer change as $T$ is lowered further. We use the results to determine the best parameters in effective models, a stiffness in a height model, an effective field theory; as well as a further-neighbour coupling in an extended Potts model. In particular the latter can then be simulated for over 10$^6$ spins, which enables us to verify the critical behavior encapsulated by the height model. Thus we identify which ordering scenario applies: we obtain spin ordering with a remarkably small ordered moment, which we estimate to be about an order of magnitude below fully developed order. In the remainder of this manuscript, we give an account of our analysis, starting with a review of the unusual behaviour of the kagome magnet as the temperature is lowered.


\ssection{Complex energy landscape}      As the kagome magnet is cooled from high temperatures, it develops short-range order like any other magnet. Upon cooling further, its frustrated nature asserts itself: at the Curie temperature, $J$, there is no sign of any ordering predicted by mean-field theory, despite the fact that the spins in each triangle firmly point at 120$^\circ$ to one another. Instead, it is not until much lower $T\alt10^{-3}J$ that all spins adopt a common plane 
        
        This coplanar ordering is described mathematically by two order parameters \cite{Zhito}: a {\em quadrupolar} (also known as nematic or coplanar) one for the direction normal to the plane; and an octupolar one as follows. If one denotes the angle the spins make in the plane with respect to a reference spin by $\theta$, it is $\exp(3 i \theta)$ which orders. Crucially, the existence of dipolar (spin) order, in  $\exp(i \theta)$, remains an open question.       
        
It should be noted that since the kagome lattice is two-dimensional, di-, quadru- and octupolar orders in fact only set in algebraically, being cut off on a lengthscale $\xi_{MW}$ for $T>0$, below which scale our analysis applies. Crucially, however, $\xi_{MW}$ diverges {\em exponentially} as the temperature is lowered, so that the ordered regimes are still well-defined in practise even at small but nonzero $T$.

As the ground state is continuously degenerate, the energy landscape has many zero-energy directions (flat valleys). However, at non-zero temperature, in the free energy landscape -- which incorporates entropic effects -- the flat valley are replaced by robust local minima at each coplanar state.  The coplanar states are discrete, and can be mapped onto the ground states of the antiferromagnetic Potts model on the kagome lattice (which is in turn equivalent to the three-colouring model of the honeycomb lattice): each spin points in one of three directions at mutual angles of 120$^\circ$ with one another. Labelling these with the colours red, green and blue, the coplanar states are those where each triangle hosts a spin of each colour. Below, we will use the fact that  these discrete models are exactly soluble \cite{BaxterPotts}, e.g.\ it is known that there are exponentially many coplanar states, $N_{c}\sim 1.13^{N}$, where $N$ is the total number of spins.

Unusually, the origin of the coplanar ordering is entropic -- coplanar states have particularly soft fluctuations around them: linear spin-wave analysis finds an {\em identical spectrum} for all of them, including an entire band of zero-energy excitations \cite{CHS}. It is thus only when anharmonic fluctuations are taken into account that the different coplanar states acquire different free energies. No algorithm has so far managed to probe these free energy differences, due to the impossibility of surmounting the entropic barriers between coplanar ground states sufficiently efficiently. Here we present an algorithm which achieves this as follows.

\ssection{Numerical algorithm} At low temperature, fluctuations around an ideal coplanar (Potts) state are small. We can thus identify a Potts configuration nearest to the state of the system. We then rotate spins on an alternating loop of spins of two colours so that these two colours are interchanged. Crucially, we have found a way of mapping configurations into each other which are equivalent as far as the linearised equations of motion are concerned -- whether or not to accept a move thus depends only on precisely the {\em anharmonic} terms responsible for the low-temperature state selection. Our algorithm implements the semiclassical dynamical symmetry identified in Ref.~\onlinecite{HasMoe_95}. 

Firstly, we construct the Potts configuration by assigning color  R to all spins with a positive projection onto a randomly chosen initial spin. Then, the other spins along the resulting loops are alternatingly colored  G--B. Next, for each loop, we evaluate the average orientations $\pmb{\Upsilon}_\alpha$ of spins colored  $\alpha =$R, G or B. Finally, we attempt a move consisting of rotating spins G (B) by the angle $\varangle$ ($\pmb{\Upsilon}_B$, $\pmb{\Upsilon}_G$) in opposite directions in the nematic plane of the loop; this move is accepted by a standard Metropolis condition. To see how the dynamical symmetry is implemented, write a spin ${\bf S}_{\alpha, i} = \pmb{\Upsilon}_\alpha + {\bf s}_{\alpha,i}$, so that the total spin of a triangle $i$, $\pmb{\ell}_i = \sum_\alpha {\bf S}_{\alpha, i} = \sum_\alpha \pmb{\Upsilon}_\alpha + \sum_\alpha  {\bf s}_{\alpha, i}$. At the lowest temperatures, $\sum_\alpha \pmb{\Upsilon}_\alpha = 0$ (the ground state condition), and the rotation operation ${\cal R}_R \pmb{\Upsilon}_R = \pmb{\Upsilon}_R$, ${\cal R}_B \pmb{\Upsilon}_B = \pmb{\Upsilon}_G$ and ${\cal R}_G \pmb{\Upsilon}_G = \pmb{\Upsilon}_B$ as $\varangle (\pmb{\Upsilon}_B, \pmb{\Upsilon}_G) = \frac{2 \pi}{3}$. Now, ${\cal R}_R {\bf s}_{R,i} + {\cal R}_B {\bf s}_{B,i} + {\cal R}_G {\bf s}_{G,i} = \left( \ell^x_i, \ell^y_i, -\ell^z_i\right)$.
The linear equations of motion are invariant under this change \cite{HasMoe_95}. We can thus fully equilibrate systems with, depending on boundary conditions, up to 2025 spins at temperatures down to $T=10^{-5}J$. Lower temperatures are achievable but, as we discuss next, this is not necessary. In all the data we show here, we have imposed a stringent  condition for equilibration, namely that memory of the starting configuration be lost, even for initialisation in maximally topologically distinct sectors --- $q = 0$ and $q = \sqrt{3} \times \sqrt{3}$ \cite{KonHenSLWC}.

\begin{figure}[ht]
\begin{center}
\includegraphics[width=0.98\columnwidth]{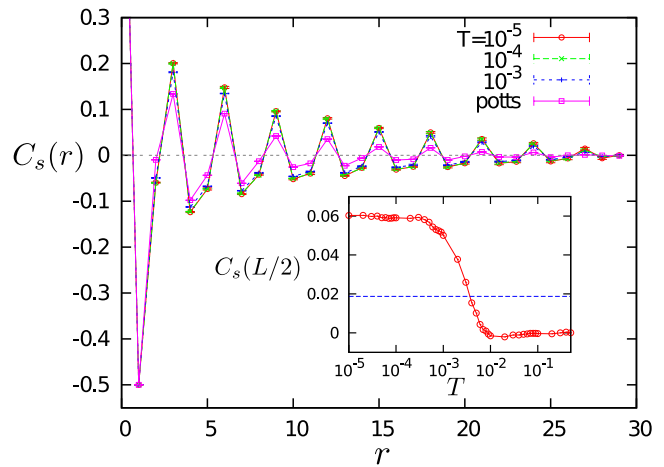}
\end{center}
\caption{
\label{fig:spincorr} 
Spin correlation functions $C_s(r) = \langle \mathbf S(r)\cdot\mathbf S(0)\rangle$ along the nearest-neighbor
directions at various temperatures. Also shown for comparison is the case of Potts model. The simulations were
done on a lattice with $N \approx 9\times 15^2$ spins with open boundary conditions. Inset: the correlations
acquire a stable low-temperature value below $T=10^{-3}J$. 
}
\end{figure}

\ssection{Correlations at low temperature} There is a limiting low-temperature regime where the correlations cease to change as $T$ is lowered further below $T\sim10^{-3}J$. This is shown in Fig. \ref{fig:spincorr} for spin-correlations in real space for a system of $N=2025$ spins with open boundary conditions.    This implies that the effective weights differentiating between different coplanar states become temperature independent, as follows. The fluctuations selecting an ordered state are subdominant compared to those which select the nematic order: the modes which cost no energy are dressed by the finite-energy ones when anharmonic effects are taken into account. To lowest order, cubic (quartic) terms appear to second (first) order in the corresponding perturbation theory. The corresponding contributions are suppressed by a factor of $T$ compared to the leading ones \cite{Henley}. Temperature therefore drops out of the resulting effective weights.  The low-temperature limit is therefore smooth. Our numerics establishes that this holds for the full problem, and not just to leading order in a perturbation theory which may or may not converge on account of the large number of zero modes. We will use this property extensively below.
        
\begin{figure}[ht]
\begin{center}
\includegraphics[width=0.9\columnwidth]{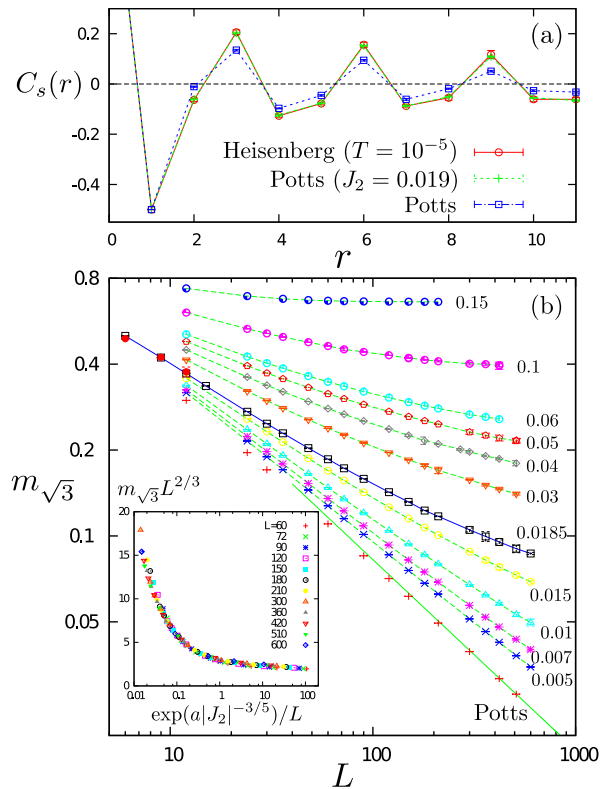}
\end{center}
\caption{
\label{fig:corrcmp} 
(a) Spin correlations of the Heisenberg model at $T=10^{-5}$ and the extended Potts model with $J_2 = 0.019$ on a lattice  with $396$ spins. 
(b) Order parameter $m_{\sqrt{3}}$ as function of linear system size $L$ for different values of
$J_2$. The solid red points for small systems are from the Heisenberg simulations. 
Inset: scaling collapse of data for $L\geq60$.
}
\end{figure}

        \ssection{Enhanced correlations at low $T$} In Fig. \ref{fig:corrcmp}, we present the correlators necessary for the demonstration of order in this regime of low temperatures. We focus on the order parameter corresponding to a tripling of the unit cell, $m_{\sqrt{3}}$, corresponding to the right state depicted in the phase diagram Fig.~\ref{fig:kagbasics}.
                 Shown for comparison are the correlations of the (unweighted) average over the coplanar Potts states, which do not take into account the anharmonic fluctuations responsible for the low-temperature regime  under consideration. The $m_{\sqrt{3}}$ deviates {\em upwards} from the decay of the Potts correlations, while the decay of $m_0$, corresponding to the left state in Fig.~\ref{fig:kagbasics}, is even faster than at higher temperatures (not shown). Next, we analyse this data in detail with the aid of effective field theories and an effective Hamiltonian.

\ssection{Height model analysis}
Historically, the question about the nature of the ordering has sometimes been phrased as an alternative between ordering in either $q=0$ or in a {$\sqrt{3}\times\sqrt{3}$} structure as these are frequent outcomes of perturbations to the classical Heisenberg model. However, the actual distinction has to be made between a power-law correlated state and an ordered state far away from the $q=0$ state, as suggested by the phase diagram in Fig.~\ref{fig:kagbasics}. This was noted long ago by Huse and Rutenberg \cite{HRkag}, based on Baxter's observation that the Potts model, i.e.\ the average over coplanar states without entropic weighting, is {\em at a critical point} in the limit of $T\rightarrow0$ \cite{BaxterPotts}. To settle the question whether the kagome magnet is in turn ordered, one therefore needs to establish whether the appropriate tuning parameter -- the stiffness of a so-called height field \cite{KonHenSLWC} -- increases rather than decreasing or remaining the same: for decreasing stiffness, the system would move away from the critical point into the KT phase, whereas for an increase, ordering ensues. 
            
\begin{figure}[ht]
\begin{center}
\includegraphics[width=0.8\columnwidth]{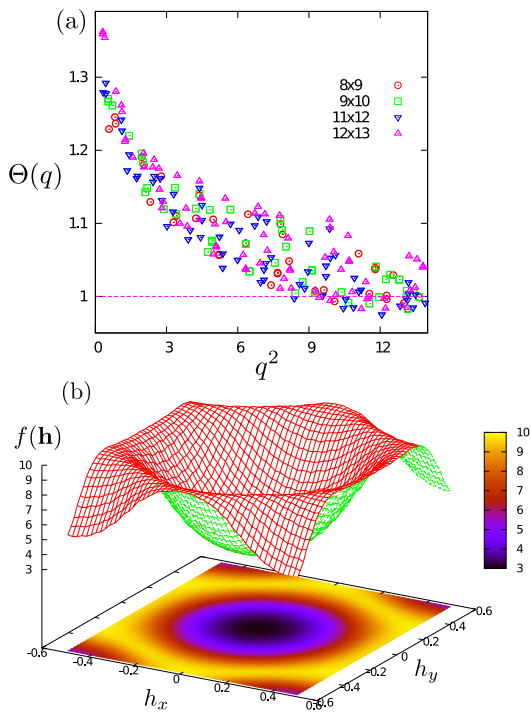}
\end{center}
\caption{
\label{fig:Rq-heis-potts}~(a)~The function $\Theta(q) \equiv \langle |\mathbf h_{\rm potts}(q)|^2 \rangle/\langle |\mathbf h_{\rm Heisenberg}(q)|^2\rangle$ denotes the ratio of the averaged height fields. A ratio $\Theta(q)>1$ for $q^2\rightarrow 0$ reflects the increased stiffness of the Heisenberg model compared to the Potts model, and the resultant ordering at low temperature. (b) Relative frequency of the averaged height vector for Heisenberg spins at $T=10^{-4}J$ for a lattice with $729$ spins and open boundary conditions. The maxima correspond to $\sqrt{3} \times \sqrt{3}$ order.
}
\end{figure}

            Given our algorithm can access the low-temperature correlations, we readily find that the stiffness has increased. This is evident already in the real-space correlations, which are significantly enhanced for Heisenberg over Potts (Fig.~\ref{fig:spincorr}). More directly the stiffness can be extracted by considering long-wavelength fluctuations of the height field $\mathbf h({\bf r})$ defined at each hexagon of the lattice. The height difference between two neighboring hexagons (labelled by $I,J$) is $\mathbf h_I - \mathbf h_J = \bm\Upsilon_{IJ}$, where ${IJ}$ denotes the site shared by the two hexagons~\cite{KonHenSLWC}. The ratio of the stiffnesses is given by $\lim_{q\rightarrow0} \Theta(q^2)\equiv \langle |\mathbf h_{\rm potts}(q)|^2 \rangle/\langle |\mathbf h_{\rm Heisenberg}(q)|^2\rangle$, as shown in Fig.~\ref{fig:Rq-heis-potts}(a). This quantity is clearly bigger than one,  and does not appear to reach a limiting value, signaling not only enhanced correlations but indeed the onset of ordering. The nature of the ordering is indicated by the histogram of the height vector, which is a periodic variable and hence best plotted over its height-space unit cell. This is what's shown in Fig.~\ref{fig:Rq-heis-potts}(b) \cite{KonHenSLWC}. The maximum (minimum) at the cell corner (centre) point corresponds to the $q=\sqrt{3}\times\sqrt{3}$ $(q=0)$ state, respectively, and the difference between these extrema is greatly enhanced over the unweighted Potts model (not shown).

            This establishes that the Heisenberg antiferromagnet is {\em more ordered} than the Potts model which describes the intermediate, coplanar, temperature regime: the magnet finds itself on the ordered side of the transition.

        
\ssection{Effective Hamiltonian}
The weakness of the resulting order turns out to be a blessing in disguise: it allows us to use an effective low-temperature Hamiltonian for the Heisenberg magnet which is only a {\em small} perturbation of the antiferromagnetic Potts model. 
This is demonstrated in Fig.~\ref{fig:corrcmp}(a), where we find quantitative agreement for the spin correlations {at all distances} between Heisenberg and an extended Potts model obtained by merely adding a weak, $J_2=0.019$, second-neighbour ferromagnetic coupling. This interaction is normalized such that (un)equal Potts spins (lose) gain an `entropy' $J_2$, i.e.\ the concomitant Boltzmann factors are of the temperature-independent form $\exp(\pm J_2)$, as explained above. Of course, there may be small additional terms which we cannot accurately fit to the data of the present simulations but the agreement between the correlations from the two models is remarkable given the simplicity of the fit, and we believe this is a basis  for a good semi-quantitative approximation scheme.

With this in hand, we can now simulate the effective Potts model to get a better handle on very large system sizes. Fig.~\ref{fig:corrcmp}(b) displays the size of the $m_{\sqrt{3}}$ for different values of $J_2$ for systems of up to over $N=10^6$ sites. While $m_{\sqrt{3}}$ does not decay to an asymptotic constant      for even those large systems, it never approaches a limiting algebraic behavior either and instead always bends upwards. Even though ordering is not directly detectable in an unambiguous fashion, we can use information from the exact solution, which leads us to expect a transition at $J_2=0$. Assuming a scaling form, ${\cal F}$, appropriate for a Kosterlitz-Thouless transition \cite{KorshKag}, 
 \begin{equation}
 m_{\sqrt3}\, L^ {2/3}={\cal F}\bigl(\xi(J_2)/L\bigr), \nonumber
 \end{equation} 
we obtain  data collapse for $m_{\sqrt {3}}$, Fig.~\ref{fig:corrcmp}(b), where $\xi = \exp(a |J_2|^{-3/5})$ is the correlation length and $a \approx 0.385$ is the only fitting parameter.  In the thermodynamic limit, $m_{\sqrt{3}} \sim \mathcal{A}\, \xi^{-2/3}$. Estimating the constant $\mathcal{A}$ from large $|J_2|$, where $m_{\sqrt3}$ can be directly read off, we obtain a small but non-zero ordered moment about an order of magnitude below the full moment.



\ssection{Conclusions and outlook}
We have analyzed the asymptotic low-temperature regime of the classical kagome Heisenberg antiferromagnet with a combination of analytical and numerical approaches. We find dipolar spin order with a tripled unit cell and a very small ordered moment. It will be interesting to see how this ordering disappears as the temperature is raised. More broadly, it is desirable to study the effective model in considerably more detail as a function of $J_2$, in particular with view to the question of whether there are any additional transitions, e.g.\ involving  the chirality \cite{castelchamon}, the other marginal operator of the critical Potts model \cite{BaxterPotts}.   Additional interesting models to examine would be the ordering of analogous magnets on the pyrochlore lattice~\cite{MoeCha}, for some work on this see Ref.~\cite{MotomeClus}, or the fate of the hyperkagome magnet in the limit of the lowest temperatures \cite{Zhito}.       

Overall, this kagome magnet provides  a most striking instance of a classical system with nonlinear-fluctuation--induced order {\em very much smaller} than the maximal possible moment, which is rather unusual for classical systems. These tend to be characterized by robust order, such as the saturated ground state order of the frustrated triangular lattice Heisenberg antiferromagnet. Like in the case of the kagome $S=1/2$ quantum magnet, our classical phase is also extremely fragile,  destabilized already by additional interactions of strength of only a few percent of the leading one. 

\ssection{Note added} As we were concluding this manuscript, Ref.~\onlinecite{kaglan} appeared, which also considers the low-temperature behavior of the classical kagome Heisenberg magnet. Their main interest is the dynamics but they do also present an algorithm accessing the low-temperature regime. 

\vspace{.5cm}

\ssection{Acknowledgments}
We are grateful to Claudio Castelnovo, John Chalker, Chris Henley, David Huse, Jesper Jacobsen, Jane Kondev, Andreas L\"auchli, Shivaji Sondhi, Mathieu Taillefumier  and Mike Zhitomirsky for useful discussions.


\end{document}